\RequirePackage{fix-cm}
\documentclass[smallcondensed]{svjour3}     % onecolumn (ditto)
\smartqed  % flush right qed marks, e.g. at end of proof
\usepackage{graphicx}
\usepackage{mathptmx}      % use Times fonts if available on your TeX system
\usepackage{mathptmx}
\usepackage{amsmath}
\usepackage{tikz}
\usepackage{caption}
\usepackage{booktabs}
\usepackage{multirow}
\usepackage{soul}
\usepackage[numbers]{natbib}
\usetikzlibrary{arrows}

\usepackage{algorithm,algpseudocode}
\newtheorem{exmp}{Example}
\algnewcommand\algorithmicinput{\textbf{Input:}}
\algnewcommand\Input{\item[\algorithmicinput]}
\algnewcommand\algorithmicoutput{\textbf{Output:}}
\algnewcommand\Output{\item[\algorithmicoutput]}
\algnewcommand\algorithmicforeach{\textbf{for each}}
\algdef{S}[FOR]{ForEach}[1]{\algorithmicforeach\ #1\ \algorithmicdo}

\begin{document}

\title{Heuristic Based Induction of Answer Set Programs}
\subtitle{From Default theories to combinatorial problems}


\author{Farhad Shakerin         \and
        Gopal Gupta %etc.
}


\institute{F. Shakerin \at
              The University of Texas at Dallas \\
              \email{fxs130430@utdallas.edu}           %  \\
           \and
            G. Gupta \at
              The University of Texas at Dallas \\
              \email{gupta@utdallas.edu}}

\date{Received: date / Accepted: date}

\maketitle

\begin{abstract}
Significant research has been conducted in recent years to extend
Inductive Logic Programming (ILP) methods to induce Answer Set Programs (ASP). 
These methods perform an exhaustive search for the correct 
hypothesis by encoding an ILP problem instance as an ASP program. 
Exhaustive search, however, results in loss of scalability. In addition, the language bias employed
in these methods is overly  restrictive too. In this paper we extend our previous
work on learning stratified answer set programs that have a single stable model
to learning arbitrary (i.e., non-stratified) ones with multiple stable models.
Our extended algorithm is a greedy FOIL-like algorithm, capable of inducing 
non-monotonic logic programs, examples of which includes programs for 
combinatorial problems such as graph-coloring and N-queens. To the best of our 
knowledge, this is the first heuristic-based ILP algorithm to induce answer set programs
with multiple stable models.      

\end{abstract}

\section{Introduction}

Statistical machine learning methods produce models that are not comprehensible for
humans because they are algebraic solutions to optimization problems such as 
risk minimization or data likelihood maximization. These methods do not produce 
any intuitive description of the learned model. Lack of intuitive descriptions 
makes it hard for users to understand and verify the underlying rules that
govern the model. Also, these methods cannot produce a justification for 
a prediction they compute for a new data sample. Additionally, if prior 
knowledge (background knowledge) is extended in these methods, then the
entire model needs to be re-learned. Finally, no distinction is made between 
exceptions and noisy data in these methods.

Inductive Logic Programming \cite{Muggleton91}, however, is one technique where 
the learned model is in the form of logic programming rules (Horn clauses) that 
are comprehensible to humans. It allows the background knowledge to be incrementally 
extended without requiring the entire model to be relearned. Meanwhile, the 
comprehensibility of symbolic rules makes it easier for users to understand and 
verify induced models and even edit them.

ILP learns theories in the form of Horn clause logic programs. Extending
Horn clauses with negation as failure (NAF) results in more powerful applications
becoming possible as inferences can be made even in absence of information. This 
extension of Horn clauses with NAF where the meaning is computed using the stable 
model semantics \cite{GelfondL88}---called Answer Set Programming\footnote{We use the
term answer set programming in a generic sense to refer to normal logic programs,
i.e., logic programs extended with NAF, whose semantics is given in terms of
stable models \cite{baral}.}---has many powerful applications. 
Generalizing ILP to learning answer set programs 
also makes ILP more powerful. 
For a complete discussion on the necessity of NAF in ILP, we refer the reader 
to \cite{sakama05}. 

Once NAF semantics is allowed into ILP systems, they 
should be able to deal with multiple stable models which arise due 
to presence of mutually recursive rules involving negation (called 
{\it even cycles}) \cite{baral} such as: 

{\tt 

p :- not q.

q :- not p.

}

%GG: rework
Inducing answer set programs in presence of even cycles in the background knowledge has 
first been explored in \cite{stableilp}, where the author describes the added 
expressiveness that results once background knowledge is allowed to have multiple stable 
models. Work by Otero \cite{otero} on induction of stable models formalizes induction 
of answer set programs with stable model semantics \cite{GelfondL88} such that in situations 
where $B \cup H$ ($B$ represents the background knowledge and $H$ the hypothesis) 
has multiple stable models, it is just necessary to guarantee that 
each positive example is true in at least one stable model of $B \cup H$. It also 
attempts to characterize inducing answer set programs from partial answer sets of 
$B \cup H$ (the author calls them non-complete set of examples). These partial answer sets
are treated as examples in the ILP problem. Otero also suggests that researchers should
focus on learning answer set programs that model combinatorial and planning problems, but
does not present any solution.  Addressing the problem of learning such programs 
is the goal of our research presented in this paper.

In \cite{sakama05}, Sakama introduces algorithms to induce 
a {\it categorical} logic program\footnote{A categorical logic program is an
answer set program with at most one stable model.} given the answer set of 
the background knowledge and {\it either} positive {\it or} negative examples. Essentially,
given a single answer set, Sakama tries to induce a program that
has that answer set as a stable model.
%GG: define categorical
In \cite{brave}, Sakama extends his work to learn from   
multiple answer sets. He introduces {\it brave} induction, where the learned 
hypothesis $H$ is such that {\it some} of the answer sets of 
$B \cup H$ cover the positive examples. The limitation of this work is that it accepts only 
one positive example as a conjunction of atoms. It does not take into account negative examples at all.
%GG: fix above sentence
Cautious induction, the counterpart of brave induction, is also 
too restricted as it can only induce atoms in the intersection 
of all stable models. Thus, neither brave induction nor cautious induction are able 
to express situations where something should hold in all 
or none of the stable models. An example of this limitation arises in the graph
coloring problem where the following should hold in all answer sets: no two neighboring 
nodes in a graph should be painted the same color.

ASPAL \cite{DBLP:conf/ilp/CorapiRL12} is the first ILP system to learn answer 
set programs by encoding ILP problems as ASP programs and having an ASP solver 
find the hypothesis. Its successor ILASP \cite{ILASP}, is an ILP system capable of 
inducing hypotheses expressed as answer set programs too. ILASP defines a framework that subsumes 
brave/cautious induction and allows much broader class of problems relating to learning answer
set programs to be handled by ILP.  However, the algorithm exhaustively searches the space of 
possible clauses to find one that is consistent with all examples and 
background knowledge. To make this search feasible, it prohibits predicate 
invention, i.e., learning predicates other than the target predicate(s). Resorting to
exhaustive search and not allowing predicate invention are 
weaknesses of ILASP that limit its applicability to many useful situations. Our
research presented in this paper does not suffer from these problems.

XHAIL \cite{XHAIL} is another ILP system capable of learning non-monotonic 
logic programs. It heavily incorporates abductive logic programming to search 
for hypotheses. It uses a similar language-bias as ILASP does, and thus suffers
from the limitations similar to ILASP. It also does not support the notion of inducing 
answer set programs from partial answer sets.

All the systems discussed above, resort to an exhaustive search for the hypothesis.
In contrast, traditional ILP systems (that only learn Horn clauses),
 use heuristics to guide their search. Use of heuristics allows these system to avoid
an exhaustive search. These system usually start with the most general clauses and then 
specialize them. They are better suited for large-scale data-sets with noise, 
since the search can be easily guided by heuristics. 
FOIL \cite{Quinlan90} is a representative of such algorithms. However, handling negation in FOIL is somewhat 
problematic as we will soon show. Also, FOIL can not handle background knowledge 
with multiple stable models, nor it can induce answer set programs.

Recently we developed an algorithm called FOLD \cite{fold} to automate 
inductive learning of default theories represented as stratified answer
set programs. FOLD (First Order Learner of Default rules) extends the 
FOIL algorithm and is able to learn answer set programs that represent the underlying knowledge
very succinctly. 
However, FOLD is only limited to dealing with stratified answer set programs, i.e.,
mutually recursive rules through negation are not allowed in the background knowledge
or the hypothesis.  Thus, FOLD is incapable of handling cases where the
background knowledge or the hypotheses admits multiple stable models. In this paper, 
we extend the FOLD algorithm to allow both the background
knowledge and the hypothesis to have multiple stable models. 
The extended FOLD algorithm---called the XFOLD algorithm---is much more general than previously
proposed methods. 

This paper makes the following novel contributions: it presents the XFOLD algorithm,
an extension of our previous FOLD algorithm, that can
handle background knowledge with multiple stable models as well as allow 
inducing of hypotheses that have multiple stable models. To the best of our knowledge, XFOLD is the first 
heuristic based algorithm to induce such hypotheses. The XFOLD algorithm
can learn ASP programs to solve combinatorial problems such as graph-coloring and N-queens.
Because the XFOLD algorithm is based on heuristic search, it is also scalable. Lack of
scalability is a major problem in previous approaches.

The rest of this paper is organized as follows: In section 
\ref{sec:background}, we present the motivation of the FOLD algorithm by 
recalling some of the problems in FOIL algorithm. In section \ref{sec:fold}, we 
introduce the FOLD algorithm. In section \ref{sec:foldasp}, we present our 
extension to the FOLD algorithm, called XFOLD, to induce answer set programs with multiple 
stable models. In section \ref{sec:application}, we show how XFOLD algorithm can 
induce programs for solving combinatorial problems. In section \ref{sec:related}, we
present related work while in 
section \ref{sec:conclusion}, we present our conclusions and future work.   

We assume that the reader is familiar with answer set programming and stable
model semantics. Books by Baral \cite{baral} and Gelfond and Kahl \cite{gelfondbook}
are good sources of background material.

\section{Background}
\label{sec:background}

In this section we describe our work on learning stratified answer set programs, i.e.,
learning hypothesis without cyclical rules using background knowledge that also does not
have cyclical rules. The learning algorithm, called FOLD (First Order Learning of Default 
rules) \cite{fold}, is itself an extension of the well known FOIL algorithm.
FOIL is a top-down ILP algorithm which follows a 
\textit{sequential covering} approach to induce a hypothesis. The FOIL 
algorithm is summarized in Algorithm \ref{algo:foil}. This algorithm repeatedly 
searches for clauses that score best with respect to a subset of 
positive and negative examples, a current hypothesis and a heuristic called 
\textit{information gain} (IG). The FOIL algorithm learns a target predicate
that has to be specified. Essentially, the target predicate appears as the head of 
the learned goal clause that FOIL aims to learn.

\begin{algorithm}
\caption{Overview of the FOIL algorithm}
\label{algo:foil}
\begin{algorithmic}[1]
\Input $goal,B,E^+,E^-$ 
\Output Hypothesis H \\
Initialize $H \gets \emptyset $
\While{\textbf{not}($stopping \ criterion$)}
	\State $c \gets$ \texttt{\{goal :- true.\}}
	\While{\textbf{not}($stopping \ criterion$)}
		\For{all $ \ c' \in \rho (c)$}
        	\State $compute \ score(E^+,E^-,H \cup \{c'\},B)$
    	\EndFor
    	\State let $\hat{c}$ be the $c' \in \rho(c)$ with the best score  
    	\State $c \gets \hat{c}$
    \EndWhile	
    \State add $\hat{c}$ to $H$
    \State $E^+ \gets E^+ \setminus covers(\hat{c},E^+,B)$
    
\EndWhile 
\end{algorithmic}
\end{algorithm}
The inner loop searches for a clause with the highest information gain using a 
general-to-specific hill-climbing search. To specialize a given clause $c$, a 
refinement operator $\rho$ under $\theta$-subsumption \cite{plotkin70} is 
employed. The most general clause is \texttt{\{p($X_1,...,X_n$) :- true.\}}, 
where the predicate \texttt{p/n} is the target and each $X_i$ is a variable. 
The refinement operator specializes the current clause \texttt{\{h :- 
$b_1$,...,$b_n$.\}}. This is realized by adding a new literal \texttt{l} to the 
clause, which  yields the following: \texttt{\{h :- $b_1$,...,$b_n$,l\}}. The 
heuristic based search uses  \textit{information gain}. In FOIL, information 
gain for a given clause is calculated as follows 
\cite{Mitchell97}:
\begin{equation}
IG(L,R) = t\left(log_2 \frac{p_1}{p_1 + n_1} - log_2 \frac{p_0}{p_0+ n_0} 
\right)
\end{equation}
where $L$ is the candidate literal to add to rule $R$, $p_0$ is the number of 
positive bindings of $R$, $n_0$ is the number of negative bindings of $R$, 
$p_1$ is the number of positive bindings of $R+L$, $n_1$ is the number of 
negative bindings of $R+L$, $t$ is the number of positive bindings of $R$ also 
covered by $R+L$.

FOIL handles negated literals in a naive way by adding the literal $not \ L$ to 
the set of specialization candidate literals for any existing candidate $L$. 
This approach leads to learning predicates that do not capture the concept 
accurately as shown in the following example:
\begin{exmp}
\label{ex:pinguin}
$ B, E^+$ are background knowledge and positive examples 
respectively under \textit{Closed World Assumption}, and the target predicate 
is \texttt{fly}.
\end{exmp}
\begin{center}
\begin{tabular}{rll}
$B:$  & \texttt{bird(X) :- penguin(X).} & \\
      & \texttt{bird(tweety).}          & \texttt{bird(et).} \\
      & \texttt{cat(kitty).}            & \texttt{penguin(polly).} \\
$E^+:$& \texttt{fly(tweety).}           & \texttt{fly(et).} \\
\end{tabular}
\end{center}
The FOIL algorithm would learn the following rule:
\begin{center}
    \begin{tabular}{l}
        \texttt{fly(X) :- not cat(X), not penguin(X).}
    \end{tabular}
\end{center}
which does not yield a constructive definition, even though it covers all the 
positives (tweety is not a penguin and et is not a cat) and no negatives 
(neither cats nor penguins fly). In fact, the correct theory in this 
example is  as follows: "{\it Only birds fly but, among them there are 
exceptional ones who do not fly}". It translates to the following logic programming rule:
\begin{center}
    \begin{tabular}{l}
        \texttt{fly(X):- bird(X), not penguin(X).}
    \end{tabular}
\end{center}
which FOIL fails to discover.

\section{FOLD Algorithm}
\label{sec:fold}
The intuition behind FOLD algorithm is to learn a concept in terms of a default 
and possibly multiple exceptions (and exceptions to exceptions, and so on). 
Thus, in the bird example given above, we would
like to learn the rule that {\tt X} flies if it is a bird and not a penguin, rather
than that all non-cats and non-birds can fly. FOLD tries first to learn the 
default by specializing a general rule of the form 
\texttt{\{goal($V_1,...,V_n$) :- true.\}} with positive literals. 
As in FOIL, each specialization must rule out some already covered negative 
examples without decreasing the number of positive examples covered 
significantly. Unlike FOIL, no negative literal is used at this stage. Once 
the IG becomes zero, this process stops. At this point, if any negative 
example is still covered, they must be either noisy data or 
exceptions to the current hypothesis. Exceptions are separated from noise via 
distinguishable patterns in negative examples \cite{Srinivasan96}. In other 
words, exceptions could be learned by calling the same algorithm recursively.
This swapping of positive and negative examples, then recursively calling the
algorithm can continue, so that we can learn exceptions to exceptions, and so on. 
%Thus, FOLD swaps the current positive and negative examples and recursively calls the FOLD 
%algorithm to learn the exception rule(s). 
Each time a rule is discovered for 
exceptions, a new predicate \texttt{ab($V_1,...,V_n$)} is introduced. To avoid 
name collisions, FOLD appends a unique number at the end of the string "ab" to 
guarantee the uniqueness of invented predicates. It turns out that the outlier 
data samples are covered neither as default nor as exceptions. If outliers are
present, FOLD identifies and enumerates them to make sure that the algorithm converges.

Algorithm \ref{algo:fold} shows a high level implementation of the FOLD 
algorithm. At lines 
1-8, function FOLD, serves like the FOIL outer loop. At line 3, FOLD starts 
with 
the most general clause (e.g. \texttt{fly(X) :- true}). At line 4, this clause 
is refined by calling the function $SPECIALIZE$. At lines 5-6, set of positive 
examples and set of discovered clauses are updated to reflect the newly 
discovered clause. 

At lines 9-29, the function $SPECIALIZE$ is shown. It serves 
like the FOIL inner loop. At line 12, by calling the function 
ADD\_BEST\_LITERAL 
the ``best'' positive literal is chosen and the best IG as well as the 
corresponding clause is returned. At lines 13-24, depending on the IG value, 
either the positive literal is accepted or the EXCEPTION function is called. 
If, at the very first iteration, IG becomes zero, then a clause that just 
enumerates the positive examples is produced. A flag called $first\_iteration$ 
is used to differentiate the first iteration. At lines 26-27, the sets of 
positive and negative 
examples are updated to reflect the changes of the current clause. At line 19, 
the EXCEPTION function is called while swapping $E^+$ and $E^-$.    

At line 31, the ``best'' positive literal that covers more positive 
examples and fewer negative examples is selected. Again, note the current 
positive examples 
are really the negative examples and in the EXCEPTION function, we try to find 
the 
rule(s) governing the exception. At line 33, FOLD is recursively called to 
extract this rule(s). At line 34, a new \texttt{ab} predicate is introduced and 
at 
lines 35-36 it is associated with the body of the rule(s) found by the 
recurring FOLD function call at line 33. Finally, at line 38, default and 
exception are combined together to form a single clause.

The FOLD algorithm, once applied to Example \ref{ex:pinguin}, yields the 
following clauses:
\begin{center}
    \begin{tabular}{l}
        \texttt{fly(X):- bird(X), not ab0(X).}\\
        \texttt{ab0(X):- penguin(X).}
    \end{tabular}
\end{center}

\begin{algorithm}

\caption{FOLD Algorithm}
\label{algo:fold}
\begin{algorithmic}[1]
\Input $goal, B,E^+,E^-$ 
\Output  \Statex $D = \{c_1,...,c_n\}$ \Comment{defaults' clauses}
		 \Statex $AB = \{ab_1,...,ab_m\}$ \Comment{exceptions/abnormal 
clauses}
\Function{FOLD}{$E^+,E^-$} 
\While{($size(E^+) > 0 $)}
\State $c \gets (goal$ :- $ \ true.)$
\State $\hat{c} \gets \Call{specialize}{{c},{E^+},{{E^-}}}$
\State $E^+ \gets E^+ \setminus 
covers(\hat{c},E^+,B)$
\State $D \gets D \cup \{ \hat{c} \}$
\EndWhile 
\EndFunction
\Function{SPECIALIZE}{${c},{E^+},{E^-}$}
\State $first\_iteration \gets true$
\While{($size(E^-) > 0 $)}
\State  $(c_{def}, \hat{IG}) \gets 
\Call{add\_best\_literal}{{c},{E^+},{{E^-}}}$
\If{$\hat{IG} > 0$}
	\State $\hat{c} \gets c_{def} $
\Else
	 \If{$first\_iteration$}
			\State $\hat{c} \gets enumerate(c,E^+)$
	 \Else	
			\State $\hat{c} \gets 
\Call{exception}{{c},{E^-},{{E^+}}}$
			 \If {$\hat{c} = null$}
						\State $\hat{c} \gets 
enumerate(c,E^+)$
					\EndIf	
	 \EndIf
\EndIf
\State $first\_iteration \gets false$
\State $E^+ \gets E^+ \setminus 
covers(\hat{c},E^+,B)$
\State $E^- \gets E^- \setminus 
covers(\hat{c},E^-,B)$

\EndWhile
\EndFunction

\Function{EXCEPTION}{${c_{def}},{E^+},{E^-}$}
\State  $\hat{IG} \gets 
\Call{add\_best\_literal}{{c},{E^+},{{E^-}}}$
\If{$\hat{IG} > 0$}
	\State $ c\_set \gets \Call{FOLD}{E^+,E^-} $
	\State $ c\_ab \gets generate\_next\_ab\_predicate()$
	\ForEach {$c \in c\_set $}
		\State $AB \gets AB \cup \{ c\_ab $:-$ \ bodyof(c) \}$
	\EndFor
	\State $\hat{c} \gets (headof(c_{def}) $:-$ \ bodyof(c), 
\textbf{not}(c\_ab))$
\Else
	\State $\hat{c} \gets null$
\EndIf

\EndFunction

\end{algorithmic}
\end{algorithm}

Now, we illustrate how FOLD discovers the above set of clauses given 
$E^+ = \{tweety,et\}$ and $E^- = \{polly,kitty\}$ and the 
goal \texttt{fly(X)}. By calling FOLD, at line 2 while loop, the clause 
\texttt{\{fly(X) :- true.\}} is specialized. Inside the $SPECIALIZE$ function, 
at line 12, the 
literal \texttt{bird(X)} is selected to add to the current clause, to get the 
clause 
$\hat{c}$ = \texttt{fly(X) :- bird(X)}, which happens to have the greatest IG 
among \texttt{\{bird,penguin,cat\}}. Then, at lines 26-27 the following updates 
are 
performed: $E^+=\{\}$,\ $E^-=\{polly\}$. A negative example 
$polly$, a penguin is still covered. In the next iteration, $SPECIALIZE$ fails 
to introduce a positive literal to rule it out since the best IG in this case 
is zero. Therefore, the EXCEPTION function is called by swapping the 
$E^+$, $E^-$. Now, FOLD is recursively called to learn a 
rule for $E^+ = \{polly\}$, $E^-=\{\}$. The recursive call 
(line 33), returns \texttt{\{fly(X) :- penguin(X)\}} as the exception. At line 
34, 
a new predicate \texttt{ab0} is introduced and at lines 35-37 the clause 
\texttt{\{ab0(X) :- penguin(X)\}} is created and added to the set of invented 
abnormalities namely, AB. At line 38, the negated exception (i.e \texttt{not 
ab0(X)}) and the default rule's body (i.e \texttt{bird(X)}) are compiled 
together to 
form the clause \texttt{\{fly(X) :- bird(X),not ab0(X)\}}.     

Note, in two different cases $enumerate$ function is called: i) At very first 
iteration of specialization if IG is zero for all the positive literals. 
ii) When the $Exception$ routine fails to find a rule governing negative 
examples. Whichever is the case, corresponding samples are considered as noise. 
The following example shows a learned logic program in presence of noise. In 
particular, it shows how $enumerate$ function works: It generates clauses in 
which the variables of the goal predicate can be unified with each member of a 
list of the examples for which no pattern exists.
\begin{exmp}
Similar to Example \ref{ex:pinguin}, plus we have an extra positive example 
fly(jet) without any further information:
\end{exmp}
\begin{center}
\begin{tabular}{cll}
$B:$    & \texttt{bird(X) :- penguin(X).} & \\
                   & \texttt{bird(tweety).} & \texttt{bird(et).} \\
                   & \texttt{cat(kitty).}   & \texttt{penguin(polly).}\\
$E^+:$  & \texttt{fly(tweety).               fly(jet).}  & \texttt{fly(et).} \\
\end{tabular}
\end{center}

FOLD algorithm on the Example 4.1 yields the following clauses:
\begin{center}
    \begin{tabular}{l}
        \texttt{fly(X) :- bird(X), not ab0(X).}\\
        \texttt{fly(X) :- member(X,[jet]).}\\
        \texttt{ab0(X) :- penguin(X).}
    \end{tabular}
\end{center}
FOLD recognizes $jet$ as a noisy data. $member/2$ is a built-in logic programming predicate in 
that tests the membership of an atom in a list.

Sometimes, there are nested levels of exceptions. The following example shows 
how FOLD manages to learn the correct theory in presence of nested exceptions.
\begin{exmp}
Birds and planes normally fly, except penguins and damaged planes that can't. 
There are super penguins who can, exceptionally, fly.
\end{exmp}
\begin{center}
    \begin{tabular}{cl}
    $B:$       & \texttt{bird(X) :- penguin(X).} \\
               & \texttt{penguin(X) :- superpenguin(X).} \\
               & \texttt{bird(a).     bird(b).    penguin(c).    penguin(d). } 
\\
               & \texttt{superpenguin(e).    superpenguin(f).} \\
               & \texttt{plane(g).    plane(h).   plane(k).} \\
               & \texttt{damaged(k).  damaged(m).} \\
    $E^+:$     & \texttt{fly(a). \ \ \ fly(b). \ \ \ fly(e).}\\
               & \texttt{fly(f). \ \ \ fly(g). \ \ \ fly(h).}
    \end{tabular}
\end{center}
FOLD algorithm learns the following theory:
\begin{center}
    \begin{tabular}{l}
        \texttt{fly(X) :- plane(X), not ab0(X).}\\
        \texttt{fly(X) :- bird(X), not ab1(X).}\\
        \texttt{fly(X) :- superpenguin(X).}\\
        \texttt{ab0(X) :- damaged(X).} \\
        \texttt{ab1(X) :- penguin(X).}
    \end{tabular}
\end{center}

\noindent
Table \ref{table:accuracy}, presents our experiments with UCI benchmark datasets \cite{Lichman:2013}. In this experiment, we ran FOLD on each dataset and measured the accuracy using a 10-fold cross-validation and the results are compared against that of Aleph \cite{aleph}. Aleph is a popular ILP system that has been widely used in prior work. To induce a clause, Aleph starts by building the most specific clause, which is called the ``bottom clause", that entails a seed example. Then, it uses a branch-and-bound algorithm to perform a general-to-specific heuristic search for a subset of literals from the bottom clause to form a more general rule. In most cases, our FOLD algorithm outperforms Aleph in terms of accuracy and succinctness of induced rules.

FOLD handling of negation and numeric constraints, yields intuitive and precise results. For instance, in UCI Labor-negotiations, which is a dataset of final settlements in labor negotiations in Canadian industry, the following hypothesis is induced by FOLD:

\begin{center}
    \begin{tabular}{l}
        \texttt{good\_contract(X) :- wage\_inc\_first\_year(X,A), A > 2, not ab0(X).}\\
        \texttt{good\_contract(X) :- holidays(X,A), A > 11.}\\
        \texttt{good\_contract(X) :- health\_plan\_half\_contribution(X), pension(X).}\\
        \texttt{ab0(X) :- no\_longterm\_disability\_help(X).} \\
        \texttt{ab0(X) :- no\_pension(X).} \\
    \end{tabular}
\end{center}
This hypothesis captures the highest priorities of employees in a good contract. Without having abnormality predicates, 
the hypothesis would have contained more clauses depending on the diversity of options on long term disability support and pension, 
whereas in default theory approach, as shown in this example, instead of covering examples with multiple clauses, 
a single clause is introduced as a default rule, and irrelevant predicates are excluded by abnormality predicates.

\begin{table}
\begin{tabular}{lcccc} 
\hline
dataset & size & ALEPH accuracy(\%) & FOLD accuracy(\%) & FOLD execution time(s) \\ [0.5ex] 
\hline\hline
Credit-au & 690 & 82 & 83 & 67\\
Credit-j & 125 & 53 & 81 & 20\\
Credit-g & 1000 & 70.9 & 78 & 87\\ 
Iris & 150 & 85.9 & 95 & 1.3\\ 
Ecoli & 336 & 91 & 90 & 6.1\\ 
Bridges & 108 & 89 & 90 & 0.8\\  
Labor & 57 & 89 & 94 & 0.4\\  
Acute(1) & 34 & 100 & 100 & 0.3\\
Acute(2) & 34 & 100 & 100 & 0.3\\
Mushroom & 7724 & 100 & 100 & 11.4 \\
\hline	
\end{tabular}
\caption{FOLD evaluation on UCI benchmarks }
\label{table:accuracy}
\end{table}

\section{Induction of Answer Set Programs with Multiple Stable Models}
\label{sec:foldasp}

In the previous section we assumed that the background knowledge $B$ is a 
normal logic program with one stable model and all examples belong to 
the only stable model of $ B \cup H $. This would require the language bias not 
to allow even cycles which are responsible for generating multiple 
stable models.

In this section we extend our FOLD algorithm to learn normal logic programs 
that potentially have multiple stable models. The significance of Answer Set 
Programming paradigm is that it provides a declarative semantics under which 
each stable model is associated with one (alternative) solution to the problem 
described by the program. Typical problems of this kind are combinatorial 
problems, e.g., graph coloring and N-queens. In graph coloring, one should find 
different ways of coloring nodes of a graph without coloring two nodes connected by
an edge with the same color. N-queen is the problem of placing N queens in a chessboard of size 
$N\times N$ so that no two queens attack each other.

In order to inductively learn such programs, the ILP problem definition needs to
be revisited. In the new scenario, positive examples $e \in E^+$, may not 
hold in every model. Therefore, the ILP problem described in the background 
section would only allow learning of predicates that hold in all answer sets. This 
is too restrictive. Brave induction \cite{brave}, in contrast, allows examples to 
hold only in some stable models of $B \cup H$. However, as stated in 
\cite{ILASP} and we will show using examples, this is not enough when it comes 
to learning global constraints (i.e, rules with empty head)\footnote{Recall that in
answer set programming, a constraint is expressed as a headless rule of the
form 

{\tt :- B.} 

which states that {\tt B} must be false. A headless rule is 
really a short-form of rules of the form (called odd loops over negation \cite{baral}):

{\tt p :- B, not p.}
}. Learning global
constraints is essential because certain combinations may have to be excluded 
from {\it all} answer sets.  

When $B \cup H$ has multiple stable models, there will be some instances of target 
predicate that would hold in all, none, or some of the stable models. Brave 
induction is not able to express situations in which a predicate should hold 
in all or none of the stable models. An example is a 
graph in which node 1 is colored red. In such a case, none of node 1's 
neighbors should be colored red. If node 1 happens to have node 2 as a neighbor, brave induction is not 
able to express the fact that if the predicate \texttt{red(1)} appears in any 
stable model of $ B \cup H $, \texttt{red(2)} should not. In \cite{ILASP}, the 
authors propose a new paradigm called \text{learning from partial answer sets} 
that overcomes these limitations. We also adopt this paradigm in our work presented here. 
Next, we present our XFOLD algorithm. 

\begin{definition}
A partial interpretation E is a pair $E = \langle E^{inc},E^{exc} \rangle$ of 
sets of ground atoms called inclusions and exclusions, respectively. Let $A 
= AS(B \cup H)$ denote a stable model of $B \cup H$. $A$ {\it extends} $\langle 
E^{inc},E^{exc} \rangle$ if and only if $(E^{inc} \subseteq \ A) \wedge 
(E^{exc} \cap \ A = \emptyset) $. 
\end{definition}

\begin{exmp}
\label{exmp1}
Consider the following background knowledge about a group of friends some of 
whom are in conflict with others. The individuals in conflict will not attend a 
party together. Also, they cannot attend a party if they work at the time the
party is held. We want our ILP algorithm to discover the rule(s) that will
determine who will go to the party based on the set of partial interpretations provided.
\end{exmp}
\begin{center}
    \begin{tabular}{ll}
        $B:$           & \texttt{conflict(X,Y) :- person(X), person(Y), 
conflict(Y,X).} \\
                       & \texttt{works(X) :- person(X), not off(X).} \\
                       & \texttt{off(X) :- person(X), not works(X).} \\
                       & \texttt{person(p1).   person(p2).} 
\texttt{conflict(p1,p4).} \\
               & \texttt{person(p3).   person(p4).} \texttt{conflict(p2,p3).}
    \end{tabular}
\end{center}
Some of the partial interpretations are as follows. 
The predicates g,w,o abbreviate goesToParty,works,off respectively:\\

\noindent
$E_1 = \{ \langle g(p1),g(p2),o(p1),o(p2),w(p3),o(p4),w(p5) \rangle, \langle 
g(p3),g(p4),g(p5)  \rangle \}$\\
$E_2 = \{ \langle g(p3),g(p4),g(p5),o(p1),o(p2),o(p3),o(p4),o(p5) \rangle, 
\langle g(p1),g(p2)  \rangle \}$\\
$E_3 = \{ \langle g(p1),g(p3),g(p5),o(p1),o(p2),o(p3),w(p4),o(p5) \rangle, 
\langle g(p2),g(p4)  \rangle \}$\\
$E_4 = \{ \langle g(p2),g(p5),g(p5),w(p1),o(p2),w(p3),w(p4),o(p5) \rangle, 
\langle g(p1),g(p3),g(p4)  \rangle \}$\\

In the above example, each $E_i$ for i = 1,2,3,4 is a partial interpretation
and should be extended by at least one stable model of $ B \cup H $ for a 
learned hypothesis $ H $. For instance, let's consider the hypothesis $H_1=$ 
\texttt{\{goesToParty(X) :- off(X)\}} for learning the target predicate 
\texttt{goesToParty(X)}. By plugging the background knowledge, the non-target 
predicates in $E_1$, and the hypothesis $H_1$ into an ASP solver (CLASP
\cite{gekasc12c} in our case), the stable 
model returned by the solver would contain 
\texttt{\{goesToParty(p1),goesToParty(p2),goesToParty(p4)\}}. It does not extend $E_1$. Although, 
$E^{inc}_1 \subseteq AS(B \cup H_1)$ but $ AS(B \cup H_1) \cap E^{exc}_1 \neq 
\emptyset$. It should be noted that non-target predicates are treated as 
background knowledge upon calling ASP solver to compute the stable model of 
$B \cup H$.

\begin{definition}
An XFOLD problem is defined as a tuple $P = \langle B,L,E^+,E^-,T 
\rangle$. $B$ is a answer set program with potentially 
multiple stable models called the background knowledge. $L$ is the language-bias 
such that $L = \langle M_h,M_b \rangle$, where $M_h$ (resp. $M_b$) are called the 
\textit{head} (resp. \textit{body}) \textit{mode declarations} \cite{Muggleton1995}. 
Each mode declaration $m_h \in M_h$ (resp. $m_b \in M_b$) is a literal whose 
abstracted arguments are either variable $v$ or constant $c$. Type of a variable
is a predicate defined in B. The domain of each constant should be defined 
separately. The clause \texttt{h :- $b_1$,...,$b_n$, not $c_1$,...,not $c_m$} is in the 
search space if and only if: i) $h$ is empty; ii) $h$ is an atom compatible 
with a mode declaration in $M_h$. Hypothesis $h$ is said to be compatible with a mode 
declaration $m$ if each instance of variable in $m$ is replaced by a variable,
and every constant takes a value from the associated domain. The set of 
candidate predicates in the greedy search algorithm are selected from $M_b \cup M_h$.

The requirement of mode declarations in the XFOLD algorithm
is due to a technicality: ASP solvers, need 
to ground the program, and for that matter, programmer should ensure that every 
variable is safe. A variable in \textit{head} is \textit{safe} if it occurs in 
a positive literal of \text{body}. XFOLD adds predicates required to ensure 
safety, but to keep our examples simple, we omit safety predicates in the 
paper. $E^+$ and $E^-$ are sets of partial interpretations called positive and 
negative examples, respectively. $T \in M_h$ is the target predicate's name. 
Each XFOLD run learns a single target predicate. A hypothesis $h \in L$ is an 
inductive solution of $T$ if and only if:
\begin{enumerate}
    \item $\forall e^+ \in E^+ \exists A \in AS(B \cup H)$ such that $A$ 
extends $e^+$
    \item $\forall e^- \in E^- \not \exists A \in AS(B \cup H)$ such that $A$ 
extends $e^-$
\end{enumerate}
\end{definition} 

The above definition adopted from \cite{ILASP} subsumes brave and cautious 
induction semantics \cite{brave}. Positive examples should be extended by at 
least one stable model of $B \cup H$ (brave induction). In contrast, no stable 
model of $B \cup H$ extends negative examples (cautious induction). The 
\textit{generate and test} problems such as N-queen and graph coloring could be 
induced using our XFOLD algorithm. It suffices to use positive examples 
for learning the \textit{generate} part and negative examples for
learning the \textit{test} part.

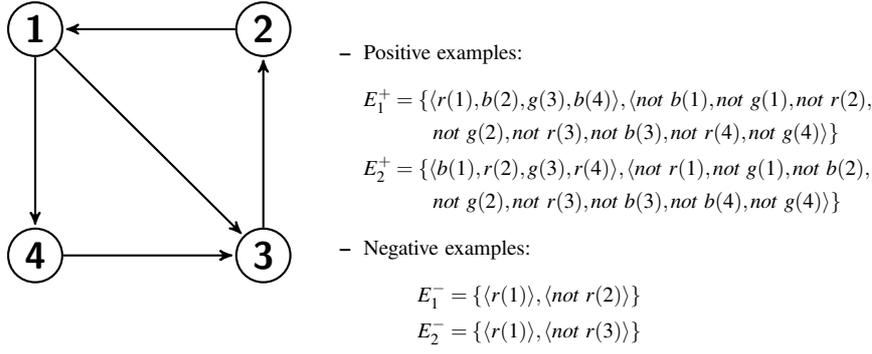
\begin{figure}[]
    \begin{tabular}{lp{0.4\textwidth}}
        \begin{tikzpicture}[baseline=0,->,>=stealth',shorten >=1pt,auto,node 
distance=3cm,
                    thick,main 
node/.style={circle,draw,font=\sffamily\Large\bfseries}]

  \node[main node] (1) {1};
  \node[main node] (2) [right of=1] {2};
  \node[main node] (3) [below of=2] {3};
  \node[main node] (4) [below of=1] {4};

  \path[every node/.style={font=\sffamily\small}]
    (1) edge node [] {} (4)
        edge node {} (3)
    (2) edge node [] {} (1)
    (3) edge node [] {} (2)
    (4) edge node [] {} (3);
\end{tikzpicture} & 
\begin{itemize}
    \item Positive examples: 
        \begin{gather*}
                E^+_1=\{  \langle r(1),b(2),g(3),b(4) \rangle,\langle not\ 
b(1),not\ g(1),not\ r(2),\\ 
      ~~~~~~~~~~~~~ not\ g(2),not\ r(3),not\ b(3),not\ r(4),not\ g(4) \rangle 
\} \\
                E^+_2=\{  \langle b(1),r(2),g(3),r(4) \rangle,\langle not\ 
r(1),not\ g(1),not\ b(2),\\ 
      ~~~~~~~~~~~~~ not\ g(2),not\ r(3),not\ b(3),not\ b(4),not\ g(4) \rangle 
\} 
        \end{gather*}
    \item Negative examples:
     \begin{gather*}
                E^-_1=\{  \langle r(1) \rangle,\langle not\ r(2)\rangle \} \\
                E^-_2=\{  \langle r(1) \rangle,\langle not\ r(3)\rangle \}
     \end{gather*}

\end{itemize}
    \end{tabular}
    \captionsetup{justification=centering}
    \caption{Partial interpretations as examples in graph coloring problem}
    \label{tab:graph-coloring}
\end{figure}

Figure \ref{tab:graph-coloring} represents the input to the XFOLD algorithm 
for learning an answer set program for graph coloring. Every positive 
example states if a node is colored red, then that node cannot be painted blue 
or green. Likewise for blue and green. However, this is not enough to learn the constraint that two nodes
connected by an edge cannot have the 
same color. To learn this constraint, negative examples are needed. For instance, 
$E^-_1$, states that if any stable model of $B \cup H$ contains 
\texttt{\{red(1)\}}, in order not to extend $E^-_1$, it should contain 
\texttt{\{not red(2)\}} or equivalently, it should not contain 
\texttt{\{red(2)\}}.

The intuition behind the XFOLD algorithm is as follows: every positive example $e$ that
is a partial interpretation is considered as a separate learning problem. A partial 
score is computed for $e$. Once all the positive examples are tested against a 
candidate clause, the overall score, i.e, the summation of all partial scores is 
stored as the score of current clause. Among all hypotheses, the one with 
highest overall score is chosen just like the single stable model case. For 
testing any given hypothesis $h$, the background knowledge $B$, all non-target 
predicates in $E^{inc}$ and the hypothesis $h$ are passed to the ASP solver as 
the input. The returned answer set is compared with the target predicates in 
$E^{inc}$ and $E^{exc}$. Next, the partial \texttt{information gain} score is 
computed. XFOLD chooses a clause with highest positive score (if one exists). 
Next, every partial interpretation is updated by removing the covered target 
predicates from $E^{inc}$ and $E^{exc}$. Once no target predicate in $E^{exc}$ 
is covered, the internal loop finishes and the discovered rule(s) are added to 
the learned theory. Just like FOLD, if no literal with positive score exists, 
swapping occurs on each remaining partial interpretation and the XFOLD algorithm 
is recursively called. In this case, instead of introducing abnormality 
predicates, the negation symbol, "-", is prefixed to the current target predicate to 
indicate that the algorithm is now trying to learn the negation of concept being 
learned. It should also be noted that swapping examples is performed slightly 
differently due to the existence of partial interpretations. The summary of 
required changes in swapping of examples is as follows:
\begin{enumerate}
    \item $\forall e \in E_{inc}$, where $e$ is and old target atom, $e$ is 
restored
    \item $\forall e \in E_{inc}$, where $e$ is and old target atom, $-e$ is 
added to $E_{exc}$
    \item $\forall e \in E_{exc}$, where $e$ is and old target atom, $-e$ is 
added to $E_{inc}$
    \item $ T \leftarrow -T$. (Target predicate T now becomes its negation, -T)   
\end{enumerate}

\begin{figure}[]
\centering
\begin{tabular}{l}
After iteration \#1: \texttt\{goesToParty(X) :- off(X)\}                               
                                                      \\ \hline
$E_1 = \{ \langle o(p1),o(p2),w(p3),o(p4),w(p5) \rangle, \langle g(p4)\rangle 
\}$                                                              \\
$E_2 = \{ \langle o(p1),o(p2),o(p3),o(p4),o(p5) \rangle, \langle 
g(p1),g(p2)\rangle \}$                                              \\
$E_3 = \{ \langle o(p1),o(p2),o(p3),w(p4),o(p5) \rangle, \langle g(p2)\rangle 
\}$                                                     \\
\st{$E_4 = \{ \langle w(p1),o(p2),w(p3),w(p4),o(p5) \rangle, \langle \rangle 
\}$}                                                          \\
After swapping $E_{inc}$,$E_{exc}$                                              
                                                      \\ \hline
$E_1 = \{\langle -g(p4),g(p1),g(p2),o(p1),o(p2),w(p3),o(p4),w(p5) \rangle, 
\langle -g(p1),-g(p2) \rangle \}$                       \\
$E_2 = \{ \langle 
-g(p1),-g(p2),g(p3),g(p4),g(p5),o(p1),o(p2),o(p3),o(p4),o(p5) \rangle, 
\langle -g(p3),-g(p4),-g(p5)\rangle \}$ \\
$E_3 = \{ \langle -g(2),g(p1),g(p3),g(p5),o(p1),o(p2),o(p3),w(p4),o(p5) 
\rangle, \langle -g(p1),-g(p3),-g(5)\rangle \}$           \\
After iteration \#1: -goesToParty(X) :- conflict(X,Y)                                 
                                                      \\ \hline
$E_1 = \{\langle g(p1),g(p2),o(p1),o(p2),w(p3),o(p4),w(p5) \rangle, \langle 
-g(p1),-g(p2) \rangle \}$                               \\
$E_2 = \{ \langle g(p3),g(p4),g(p5),o(p1),o(p2),o(p3),o(p4),o(p5) \rangle, 
\langle -g(p3),-g(p4)\rangle \}$                         \\
$E_3 = \{ \langle g(p1),g(p3),g(p5),o(p1),o(p2),o(p3),w(p4),o(p5) \rangle, 
\langle -g(p1),-g(p3)\rangle \}$                         \\
iteration \#2: -goesToParty(X) :- conflict(X,Y),goesToParty(Y)                               
                                                      \\ \hline
\st{$E_1 = \{\langle g(p1),g(p2),o(p1),o(p2),w(p3),o(p4),w(p5) \rangle, \langle 
 \rangle \}$}                                              \\
\st{$E_2 = \{ \langle g(p3),g(p4),g(p5),o(p1),o(p2),o(p3),o(p4),o(p5) \rangle, 
\langle \rangle \}$}                                        \\
\st{$E_3 = \{ \langle g(p1),g(p3),g(p5),o(p1),o(p2),o(p3),w(p4),o(p5) \rangle, 
\langle \rangle \}$}                                        \\ \hline
Hypothesis = \{ \texttt\{goesToParty(X) :- off(X), not -goesToParty(X).\}, 
\texttt\{-goesToParty(X) :- conflict(X,Y), goesToParty(Y).\}\}                      
\end{tabular}
\caption{Trace of XFOLD internal loop and recursive call on party example}
\label{tab:party-trace}
\end{figure}

Figure \ref{tab:party-trace} demonstrates execution of XFOLD for Example \ref{exmp1}. 
At the end of first iteration, the predicate \texttt{off(X)} gets the highest 
score. $E_4$ will be removed as it is already covered by the current hypothesis. 
In the second iteration, all candidate literals fail to get a positive score. 
Therefore, swapping occurs and algorithm tries to learn the predicate 
\texttt{-goesToParty(X)} as if it was an exception to the default case 
\texttt{\{goesToParty(X) :- off(X)\}}. Since the new target predicate is 
\texttt{-goesToParty(X)}, all ground atoms of \texttt{goesToParty} in $E_{inc}$ are restored 
back. The old target atoms in $E_{exc}$ are transformed to negated version 
and become members of $E_{inc}$. 

In Figure \ref{tab:party-trace}, after one iteration, $E_4$ is removed because 
all target atoms in $E_{inc}$ are already covered and targets atoms in 
$E_{exc}$ are already excluded. After swapping, XFOLD is recursively called to 
learn \texttt{-goesToParty}. After two iterations, since all examples are covered, 
the algorithm terminates.

In Example \ref{exmp1}, we haven't introduced any explicit negative example. 
Nevertheless, the algorithm was able to successfully find the cases in which 
the original target predicate does not hold (via learning \texttt{-goesToParty(X)} 
predicate). In general, it is not always feasible for the algorithm to figure 
out prohibited patterns without getting to see a very large number of positive examples.

\iffalse
It is also worth noting that \texttt{-goesToParty(X)}, in essence, 
represents classical negation. XFOLD learns the negation of 
target predicate from negative examples, but then it shifts the negated head to the body 
of the rule to produce a constraint. 
Thus, given the following rule that is learned, 

\begin{center}
    \begin{tabular}{c}
        \texttt{-goesToParty(X) :- conflict(X,Y),goesToParty(Y).}
    \end{tabular}
\end{center}

XFOLD subsequently shifts \texttt{-goesToParty(X)} from the head to the body, to turn it
into a constraint:

\begin{center}
    \begin{tabular}{c}
        \texttt{:- goesToParty(X),conflict(X,Y),goesToParty(Y).}
    \end{tabular}
\end{center}

We will show detailed examples in the applications section. 
\fi

\section{Application: Combinatorial Problems}
\label{sec:application}

A well-known methodology for declarative problem solving is the 
\textit{generate and test} methodology, whereby possible solutions to a problem 
are generated first, and then non-solutions are eliminated by testing. In 
Answer Set Programming, the \textit{generate} part is encoded by enumerating 
the possibilities by introducing even cycles. The 
\textit{test} part is realized by having constraints that would eliminate 
answer sets that violate the test conditions.
ASP syntax allows rules of the form $l\{h_1,...,h_k\}u$ such that $0 \leq l 
\leq u \leq k$ and $\forall i \in [1,k]$, $h_i \in L$, where L is the language 
bias. This is a syntactic sugar for combination of even cycles and constraints, 
which is called {\it choice rule} in the literature \cite{baral,gelfondbook}. 

ILASP \cite{ILASP} directly 
searches for choice rules by including them in the search space. XFOLD, on the 
other hand, performs the search based on $\theta$-subsumption \cite{plotkin70} 
and hence disallows search for choice rule hypotheses. Instead, it directly
learns even cycles as well as constraints. This is advantageous as it allows
for more sophisticated and flexible language bias.

It turns out that inducing the \textit{generate} part in a combinatorial 
problem such as graph-coloring requires an extra step compared to the FOLD algorithm.
For instance, \texttt{red(X)} predicate has 
the following clause:
\begin{center}
    \begin{tabular}{c}
        \texttt{red(X):- not blue(X), not green(X).}
    \end{tabular}
\end{center}
To enable XFOLD to induce such a rule, we adopted the ``Mathews Correlation 
Coefficient" (MCC) \cite{quickfoil} measure to perform the task of feature selection. MCC is 
calculated as follows:
\begin{center}
    \begin{tabular}{c}
    $ MCC = \frac{TP \times TN - FP \times 
FN}{\sqrt{(TP+FP)(TP+FN)(TN+FP)(TN+FN)}}$
    \end{tabular}
\end{center}

This measure takes into account all the four terms TP (true positive), TN (true
negative), FP (false positive) and FN (false negative) in the 
confusion matrix and is able to fairly assess the quality of classification 
even when the ratio of positive tuples to the negative tuples is not close to 
1. The MCC values range from -1 to +1. A coefficient of +1 represents a perfect 
classification, 0 represents a classification that is no better than a random classifier, and -1 
indicates total disagreement between the predicted and the actual labels. MCC 
cannot replace XFOLD heuristic score, i.e, \textit{information gain}, because 
the latter tries to maximize the coverage of positive examples, while the 
former only maximally discriminates between the positives and negatives. Nevertheless, 
for the purpose of feature extraction among the negated literals which are 
disallowed in XFOLD algorithm, MCC can be applied quite effectively. For that matter, 
before running XFOLD algorithm, the MCC score of all candidate literals are 
computed. If a predicate scores ``close" to +1, the predicate itself is added
to the language bias. If it scores ``close" to -1, its negation is 
added to the language bias. For 
example, in case of learning \texttt{red(X)}, after running the feature 
extraction on the graph given in Figure \ref{tab:graph-coloring}, XFOLD computes 
the scores -0.7, -0.5 for \texttt{green(X)} and \texttt{blue(X)}, respectively. 
Therefore, \texttt{\{not green(X),not blue(X)\}} are appended to the list of 
candidate predicates. Now, after running the XFOLD algorithm, after two 
iterations of the inner loop, it would produce the following rule:
\begin{center}
    \begin{tabular}{c}
        \texttt{red(X) :- not green(X), not blue(X).}
    \end{tabular}
\end{center}
Corresponding rules for \texttt{green(X)} and \texttt{blue(X)} are learned in a 
similar manner. This essentially takes care of the \textit{generate} part of
the combinatorial algorithm. In order 
to learn the \textit{test} part for graph coloring, we need the negative 
examples shown in Figure \ref{tab:graph-coloring}. It should be noted that in 
order to learn a constraint, we first learn a new target predicate which is the 
negation of the original one. Then we shift the negated predicate from the head to the body
inverting its sign in the process. That is, we first learn a clause of the form

{\tt -T :- b$_1$, b$_2$ $\dots$ b$_n$.}

which is then transformed into the constraint:

{\tt :- b$_1$, b$_2$ $\dots$ b$_n$, T.}

Thus, the following steps should be taken to learn constraints from negative examples:
\begin{enumerate}
    \item Add rule(s) induced for \textit{generate} part to B.
    \item $\forall e^+ \in E^+, e^- \in E^-$, if $e^-_{inc} \subseteq 
e^+_{inc}$:
        \begin{itemize}
            \item \textbf{if} $e^-_{exc}$ is of the form (\texttt{not 
$p(V_1,...V_m)$}) \textbf{then} $e^+_{inc} \leftarrow e^+_{inc} \cup \{ 
-p(V_1,...V_m)\}$ 
            \item \textbf{else} $e^+_{exc } \leftarrow e^+_{exc} \cup \{ 
-p(V_1,...V_m)\}$
        \end{itemize}
    \item compute the contrapositive form of the rule(s) learned in 
\textit{generate} part and remove the body predicates from the list of 
candidate predicates  
    \item run FOLD to learn \texttt{\-p}
    \item shift \texttt{-p} from the head to the body for each rule returned 
by FOLD
\end{enumerate}

The contrapositive of a statement has its antecedent and consequent inverted 
and flipped. For instance, the contrapositive of the clause \texttt{\{red(X) :- 
not green(X), not blue(X)\}} is shown in Figure \ref{tab:contrapositive}. 

\begin{algorithm}
\caption{Overview of the XFOLD algorithm}
\label{algo:xfold}
\begin{algorithmic}[1]
\Input $ L = \langle M_h, M_b \rangle,B,E^+,E^-$ 
\Output Hypothesis H \\

\% - \textbf{Induction of ``generate" part} - \% 
\State Initialize $H \gets \emptyset $ 
\State \textbf{let} f be new features discovered by running each $ l \in L$ and measuring MCC
\State $L \leftarrow L \cup f$
\ForEach{$t \in M_h$}
 \State $h_t \leftarrow $ FOLD $ \langle B,L,E^+_{inc},E^+_{exc},t\rangle $
 \State $H \leftarrow H \cup h_t$
\EndFor
\State $B \leftarrow B \cup H$\\

\% - \textbf{Induction of ``test" part} - \% 
\ForEach{$t \in M_h$} 
    \ForEach{ $e^+ \in E^+,e^- \in E^-$}
        \If{$e^-_{inc} \subseteq e^+_{inc}$}
            \If{$e^-_{exc}$ is of the form \texttt{not t($V_1,...,V_m$)}}
                \State $e^+_{inc} \leftarrow e^+_{inc} \cup $ \{\texttt{-t($V_1,...,V_m$)}\}
            \Else
                \State $e^+_{exc } \leftarrow e^+_{exc} \cup$ \{\texttt{-t($V_1,...,V_m$)}\}
            \EndIf
        \EndIf
    \EndFor
\EndFor
\State compute the contrapositive form for each $h \in H$ in \textit{generate} part and remove the body predicates from the list of candidate predicates $L$  
\ForEach{$t \in M_h$}
    \State $h_t \leftarrow $ FOLD $\langle B,L,E^+_{inc},E^+_{exc},-t\rangle $
    \State shift -t from the head to the body to get a constraint $\hat{h_t}$
    \State $ H \leftarrow H \cup \{ \hat{h_t} \}$
\EndFor
 
\end{algorithmic}
\end{algorithm}

\begin{figure}[]
    \centering
    \begin{tabular}{l}
        \texttt{-red(X) :- green(X).}\\
        \texttt{-red(X) :- blue(X).}
    \end{tabular}
    \caption{Contrapositive for ``generate" rule in graph-coloring}
    \label{tab:contrapositive}
\end{figure}

The reason why step 3 is necessary is the following: running FOLD without eliminating 
the literals in contrapositive rule results in learning trivial clauses shown 
in Figure \ref{tab:contrapositive}. However, as soon as those trivial choices 
are removed from search space, FOLD algorithm comes up with the next best 
hypothesis which is as follows:
\begin{center}
    \begin{tabular}{l}
        \texttt{-red(X) :- edge(X,Y), red(Y).}
    \end{tabular}
\end{center}
Shifting the predicate \texttt{-red(X)} to the body yields the following 
constraint:
\begin{center}
    \begin{tabular}{l}
        \texttt{:- red(X), edge(X,Y), red(Y).}
    \end{tabular}
\end{center}
In graph coloring problem, $M_h$ = \texttt{\{red(X), green(X), blue(X)\}}. Once 
similar examples for \texttt{green(X)} and \texttt{blue(X)} are provided, XFOLD 
is able to learn the complete solution as shown in Figure 
\ref{tab:full-graph-coloring-theory}. Algorithm \ref{algo:xfold}, presents a high level view of XFOLD to induce a \textit{generate and test} hypothesis.
\begin{figure}[]
    \centering
    \begin{tabular}{l}
        \texttt{ red(X) :- not green(X), not blue(X).}\\
        \texttt{ green(X) :- not blue(X), not red(X).}\\
        \texttt{ blue(X) :- not green(X), not red(X).}\\
         \texttt{:- red(X), edge(X,Y), red(Y).} \\
         \texttt{:- blue(X), edge(X,Y), blue(Y).} \\
         \texttt{:- green(X), edge(X,Y), green(Y).}
    \end{tabular}
    \caption{Full graph-coloring ASP theory learned by FOLD algorithm}
    \label{tab:full-graph-coloring-theory}
\end{figure}
\begin{exmp}
\label{ex:chess}

Next we discuss learning the answer set program for the 
4-queen problem: the following items are assumed: Background knowledge $B$
including predicates describing a $4 \times 4$ board, rules describing 
different ways through which two queens attack each other and  examples of the 
following form:
\begin{center}
    \begin{tabular}{ll}
     B:  & 
\texttt{attack\_r($R_1$,$C_1$,$R_2$,$C_2$):-q($R_1$,$C_1$),q($R_2$,$C_2$),$C_1 
!= C_2$, $R_1 = R_2$.} \\
         & 
\texttt{attack\_c($R_1$,$C_1$,$R_2$,$C_2$):-q($R_1$,$C_1$),q($R_2$,$C_2$),$R_1 
!= R_2$, $C_1 = C_2$.} \\
         & 
\texttt{attack\_d($R_1$,$C_1$,$R_2$,$C_2$):-q($R_1$,$C_1$),q($R_2$,$C_2$),$R_1!=
R_2$,$R_1-C_1=R_2-C_2$.}\\
         & 
\texttt{attack\_d($R_1$,$C_1$,$R_2$,$C_2$):-q($R_1$,$C_1$),q($R_2$,$C_2$),$R_1!=
R_2$,$R_1+C_1=R_2+C_2$.}\\
         \hline
     E:  & $E^+_1 = \{\langle q(2,1),q(4,2),q(1,3),q(3,4)\rangle , \langle 
q(1,1),q(1,2),...,q(4,4)\rangle \}$\\
         & ... \\
         & $E^-_1 = \{\langle q(2,1) \rangle , \langle not\ q(2,2)\rangle \}$ \\
         & $E^-_2 = \{\langle q(2,1) \rangle , \langle not\ q(2,3)\rangle \}$ \\
         & $E^-_3 = \{\langle q(4,2) \rangle , \langle not\ q(1,2)\rangle \}$ \\
         & $E^-_4 = \{\langle q(4,2) \rangle , \langle not\ q(2,3)\rangle \}$
    \end{tabular}
\end{center}
\end{exmp}
As far as the \textit{generate} part concerns, XFOLD algorithm would learn the 
following program:
\begin{center}
    \begin{tabular}{l}
        \texttt{q(X,Y) :- not -q(X,Y).}\\
        \texttt{-q(X,Y) :- not q(X,Y).}
    \end{tabular}
\end{center}
The predicate \text{-q(X,Y)} is introduced by XFOLD algorithm as a result of 
swapping the examples and calling itself recursively. After computing the contrapositive 
form, \texttt{q(X,Y), -q(X,Y)} are removed from the list of candidate 
predicates. Then based on the examples provided in Example \ref{ex:chess}, XFOLD 
would learn the following rules:
\begin{center}
    \begin{tabular}{l}
        \texttt{-q($V_1$,$V_2$) :- attack\_r($V_1$,$V_2$,$V_3$,$V_4$).}\\
        \texttt{-q($V_1$,$V_2$) :- attack\_c($V_1$,$V_2$,$V_3$,$V_4$).}\\
        \texttt{-q($V_1$,$V_2$) :- attack\_d($V_1$,$V_2$,$V_3$,$V_4$).}
    \end{tabular}
\end{center}
After shifting the predicate \texttt{-q($V_1$,$V_2$)} to the body, we get the 
following constraint:
\begin{center}
    \begin{tabular}{l}
        \texttt{:- q($V_1$,$V_2$), attack\_r($V_1$,$V_2$,$V_3$,$V_4$).}\\
        \texttt{:- q($V_1$,$V_2$), attack\_c($V_1$,$V_2$,$V_3$,$V_4$).}\\
        \texttt{:- q($V_1$,$V_2$), attack\_d($V_1$,$V_2$,$V_3$,$V_4$).}
    \end{tabular}
\end{center}
It should be noted that, since XFOLD is a sequential covering algorithm like 
FOIL, it takes three iterations before it can cover all examples which in turn 
becomes three constraints as shown above. 

\section{Related Work}
\label{sec:related}
Many researchers have tried to extend Horn ILP into richer non-monotonic logic formalisms. 
``Stable ILP" \cite{stableilp} was the first effort to explore the expressiveness of 
background knowledge with multiple stable models. A survey of extending Horn clause based ILP to 
non-monotonic logics can be found in \cite{sakama05}. In this paper Sakama also introduces algorithms to learn from the answer 
set of a categorical logic program. The algorithms learn from positive and negative 
examples separately and the approach also leads to redundant literals in the body of the 
induced clause as shown by Example \ref{sakama-issues}.

\begin{exmp}
Consider the following background knowledge and positive example:
\label{sakama-issues}
\end{exmp}
\begin{center}
\begin{tabular}{rll}
$B:$  & \texttt{bird(X) :- penguin(X).} & \\
      & \texttt{bird(tweety).}          & \texttt{bird(et).} \\
      & \texttt{bear(teddy).}           & \texttt{penguin(polly).} \\
      & \texttt{cat(kitty).}            &  \\
$E^+:$& \texttt{fly(tweety).}           & 
\end{tabular}
\end{center}
Sakama's algorithm would induce the following clause:
\begin{center}
    \begin{tabular}{l}
        \texttt{fly(X) :- bird(X), not cat(X), not penguin(X), not bear(X).}
    \end{tabular}
\end{center}

The literals \texttt{not cat(X), not bear(X)} are redundant. 
The \textit{brave} induction framework \cite{brave}, although capable of learning ASP 
programs, only admits one positive example in the form of conjunction of literals. 
As we discussed, many problems, including programs for solving combinatorial problems, 
cannot be expressed without having a notion of a negative example. ILASP \cite{ILASP}, 
introduces a framework that would allow to 
induce a hypothesis from multiple positive examples {\it bravely} (i.e., it uses brave induction), 
while it would exclude negative examples cautiously (i.e., it uses cautious induction). 
However, due to performing an exhaustive search on its predetermined language 
bias, ILASP is unable to scale up to large datasets or noisy datasets. It is not able to induce 
default theories with nested, or composite abnormality predicates to capture exceptions
as shown in Example \ref{ilasp-issues}.

\begin{exmp}
\label{ilasp-issues}
A default theory with abnormality predicate represented as conjunction of two other predicates, namely \texttt{s(X) and r(X)}. 
\end{exmp}
\begin{center}
    \begin{tabular}{l}
        \texttt{p(X) :- q(X), not ab(X).}\\
        \texttt{ab(X) :- s(X), r(X).}
    \end{tabular}
\end{center}

XHAIL \cite{XHAIL} is an ILP system capable of learning non-monotonic 
logic programs.  It relies heavily on abductive reasoning incorporated
in a three-stage algorithm. It does not support inducing from multiple partial answer sets.

\section{Conclusion and Future Work}
\label{sec:conclusion}

In this paper we presented the first heuristic-based algorithm to inductively 
learn normal logic programs with multiple stable models. The advantage of this 
work over similar ILP systems such as ILASP \cite{ILASP} is that unlike these systems,
XFOLD does not perform an exhaustive search to discover the ``best" hypothesis. 
XFOLD adopts a greedy approach,
guided by heuristics, that is scalable and noise resilient. Also, learning knowledge
patterns in terms of defaults and exceptions produces more natural and intuitive results 
that correspond to common sense reasoning employed by humans.
We also showed how our algorithm could be 
applied to induce declarative logic programs that follow the \textit{generate and 
test} paradigm for finding solutions to combinatorial problems such as graph-coloring and N-queens. 

Our XFOLD algorithm has a number of novel features absent in other prior works: 
(i) it performs a heuristic search for learning hypothesis rather than an exhaustive search
and thus is considerably more scalable; (ii) it admits predicate invention allowing us
to learn a broader class of answer set programs that cannot be learned by other systems such
as ASPAL, ILASP, and XHAIL; (iii) because of swapping of positive and negative examples,
XFOLD is able to distinguish between exceptions and noise, producing more succinct hypotheses.

There are two main avenues for future work: (i) handling large datasets 
using methods similar to QuickFoil \cite{quickfoil}. In QuickFoil, all the operations of 
FOIL are performed in a database engine. Such an implementation, along with pruning techniques and query 
optimization tricks can make the XFOLD training much faster; (ii) XFOLD learns function-free answer set programs.
We plan to investigate extending the language bias towards accommodating functions.

\begin{acknowledgements}
Authors are partially supported by NSF Grant IIS 1718945.
\end{acknowledgements}

\bibliographystyle{spbasic}      % basic style, author-year citations
\bibliography{myilp}   % name your BibTeX data base

\end{document}